\documentclass[12pt]{article}

\usepackage{authblk}
\usepackage{graphicx, setspace, amsmath, amsfonts, amssymb, fontenc,  rotating,}
\usepackage{makecell, gensymb, secdot}

\usepackage[top=1in, bottom=1in, left=0.7in, right=0.7in]{geometry}
\usepackage[utf8]{inputenc}
\usepackage{pdflscape}
\def\bp{\begin{pmatrix}}
\def\ep{\end{pmatrix}}
\def\bc{\begin{center}}
\def\ec{\end{center}}
\def\be{\begin{equation}}
\def\ee{\end{equation}}

\begin{document}
\title{Revisiting representations of quark mixing matrix}

\author{Gurjit Kaur, Aakriti Bagai, Gulsheen Ahuja$^*$ and Manmohan Gupta$^\dagger$\\Department of Physics,\\ Panjab University, Chandigarh, India.\\
\vspace{0.7cm}
* gulsheen@pu.ac.in\\
$^\dagger$ mmgupta@pu.ac.in}

\onehalfspacing
\maketitle

\begin{abstract}
Using unitarity, unlike the approaches available in the literature, we have constructed 9 independent representations of CKM matrix starting with each of the 9 elements of the matrix. The relationship of these independently constructed representations with the already available ones in the literature has been compared and discussed. Further, the implications of these representations have been explored for some of the CKM parameters such as $\delta$, J and $\epsilon_k$. Interestingly, we find that the PDG representation which is equivalent to our first representation seems to be most appropriate to incorporate the hierarchy of the elements of the CKM matrix as well as to describe the related phenomenology.

\end{abstract}
\section{Introduction}

Over the last few decades,  Cabibbo-Kobayashi-Maskawa  (CKM) \cite{ckm1,ckm2} phenomenology has registered remarkable progress on the experimental as well as theoretical front. On the experimental front, significant developments have been made in generating large amount of data for the measurement of various CKM parameters.  Several groups like Particle Data Group (PDG) \cite{pdg22}, CKMfitter \cite{ckmfit}, HFLAV \cite{hflav23}, UTfit \cite{utfit}, etc., have been actively engaged in continously updating their analyses to arrive at more and more refined conclusions.  At present, we have several CKM parameters which are determined with good deal of accuracy, e.g., the matrix elements $|V_{us}|=0.2243 \pm 0.0008$, $|V_{cb}|=(40.8 \pm 1.4) 10^{-3}$ are determined within an accuracy of a few percent \cite{pdg22}. Similarly, the angles of the unitarity triangle are also known within an error of few percent. In particular, the parameter $\sin2\beta$, representing angle $\beta$ of the unitarity triangle, is well measured with its world average being $(22.2\pm0.7)^{\degree}$ \cite{pdg22, hflav23}. Similarly, the angle $\alpha$ of the unitarity triangle is also known within
a few percent level, e.g., the world average is $(85.2^{+4.8}_{-4.3})^{\degree} $ \cite{pdg22}.

On the theoretical front, CKM paradigm  has played a crucial role in understanding several important features of flavor physics. The CKM matrix, characterised by 3 mixing angles and a CP violating phase, can have only 9 independent representations or parametrizations. In the literature \cite{jar}-\cite{beta} several representations of the CKM matrix have been discussed. In particular, Refs. \cite{jar} and \cite{fx} adopt the methodology of arriving at the representations by writing the CKM matrix as a product of three rotation matrices. In Ref. \cite{ar} attempt has been made to construct the possible representation using the unitarity constraints of the CKM matrix. Somewhat recently \cite{abg, beta}, attempt has been made to incorporate one of the angles of unitarity triangle as CP violating phase of the CKM matrix, resulting into several possible representations of CKM matrix involving 4 measurable parameters.

A closer look at the above attempts reveals that none of these emphasise clearly the fact that the given representations are the only 9 possible independent ones, nor do these explore the relation between these different representations. Also, keeping in mind the present level of measurement of the CKM parameters, it is to be noted that these attempts do not explore explicitly the usefulness of a particular representation.  It is also not clear that the recent attempt \cite{abg, beta} involving 4 directly measurable CKM parameters, including  one of the angle of unitarity triangle  in a particular representation, would lead to any advantage in carrying out the phenomenological analyses. It, therefore, becomes interesting to find 9 independent representations of the CKM matrix starting from the basic constraints of unitarity and also to check the co-relation of these with the already existing representations given by different authors. It would also be interesting to check whether any particular representation  can be preferred for carrying out particular phenomenological analysis.
 
 Keeping the above issues in mind,
  the purpose of the present paper is to construct all possible independent parametrizations of CKM matrix in rigorous and ab-initio manner. The relationship of these independently constructed representations with the already available representations in the literature would also be explored. Further, the implications of these representations, incorporating unitarity constraints, on some of the  CKM parameters would be explored using the latest data.
\section{Revisiting representations of the CKM matrix}
Before proceeding further, a brief discussion of the presently known representations is perhaps desirable. To begin with, let us define the CKM matrix, e.g., 
\be 
\begin{pmatrix}
d^\prime\\s^\prime\\b^\prime
\end{pmatrix}=V_{CKM} \begin{pmatrix}
d\\s\\b
\end{pmatrix},~ \text{where}~ V_{CKM}=\begin{pmatrix}
V_{ud}&V_{us}&V_{ub}\\
V_{cd}&V_{cs}&V_{cb}\\
V_{td}&V_{ts}&V_{tb}\\
\end{pmatrix}.
\ee
 The mixing matrix $V_{CKM}$ being a $3\times3$ unitary matrix can have only 9 independent representations.  
We first discuss the approach given by C. Jarlskog \cite{jar}, for the sake of readability as well as to facilitate discussion, we reproduce their methodology here. According to Ref. \cite{jar}, the CKM matrix can be written as a product of three rotation matrices, e.g., $ R_{12}$, $R_{23}$ and $R_{13}$, given by 
\begin{equation}
 R_{12}(\theta_{12})= \begin{pmatrix} c_{12} & s_{12} & 0\\
  -s_{12} & c_{12} & 0\\
  0&0&1 \end{pmatrix},  ~~ R_{23}(\theta_{23})= \begin{pmatrix} 1&0&0\\
  0& c_{23}& s_{23}\\
  0& -s_{23} &  c_{23} \end{pmatrix}, ~~ R_{13}(\theta_{13})= \begin{pmatrix}c_{13}& 0& s_{13}\\
  0&1&0\\
  -s_{13}& 0& c_{13} \end{pmatrix}, \label{rot} 
\end{equation}
where $s_{12}$, $s_{23}$ and $s_{13}$ denote the sines of the three mixing angles.
  The author mentions 12 different ways to arrange product of these rotation matrices, yielding
 \be 
 \begin{array}{l}
R=R_{23}(\theta_{23})R_{13}(\theta_{13})R_{12}(\theta_{12}),\hspace{1cm}
  R=R_{23}(\theta_{23})R_{12}(\theta_{12})R_{13}(\theta_{13}),\\
 R=R_{12}(\theta_{12})R_{23}(\theta_{23})R_{13}(\theta_{13}),\hspace{1cm}
 R=R_{12}(\theta_{12})R_{13}(\theta_{13})R_{23}(\theta_{23}),\\
  R=R_{13}(\theta_{13})R_{12}(\theta_{12})R_{23}(\theta_{23}),\hspace{1cm}
 R=R_{13}(\theta_{13})R_{23}(\theta_{23})R_{12}(\theta_{12}),\\
 R=R_{12}(\theta_{12})R_{23}(\theta_{23})R_{12}(\theta_{12}^\prime),\hspace{1cm}
R=R_{12}(\theta_{12})R_{13}(\theta_{13})R_{12}(\theta_{12}^\prime),\\
 R=R_{23}(\theta_{23})R_{12}(\theta_{12})R_{23}(\theta_{23}^\prime),\hspace{1cm}
 R=R_{23}(\theta_{23})R_{13}(\theta_{13})R_{23}(\theta_{23}^\prime),\\
  R=R_{13}(\theta_{13})R_{12}(\theta_{12})R_{13}(\theta_{13}^\prime),\hspace{1cm}
R=R_{13}(\theta_{13})R_{23}(\theta_{23})R_{13}(\theta_{13}^\prime),\\
 \end{array}
 \label{eq.jar}
\ee
where $\theta_{ij}^\prime \neq \theta_{ij}$.
To obtain a possible unitary representation of CKM matrix, it was suggested that phase factor $\delta$ could be added in 3 different ways leading to 36 representations of the CKM matrix, obviously all of these cannot be independent.
 Considering the possibility $R=R_{23}(\theta_{23})R_{13}(\theta_{13})R_{12}(\theta_{12})$, the phase factor $\delta$ can be added in 3 possible ways as
$R_{23}(\theta_{23},\delta)$  $ R_{13}(\theta_{13},0)$ $ R_{12}(\theta_{12},0)$ or $R_{23}(\theta_{23},0)$ $ R_{13}(\theta_{13},\delta)$ $R_{12}(\theta_{12},0)$ or $ R_{23}(\theta_{23},0)$ $R_{13}(\theta_{13},0)$ $R_{12}(\theta_{12}, \delta)$. For example,
$$\begin{scriptsize}
R_{23}(\theta_{23},\delta)= \begin{pmatrix} 1&0&0\\
  0& c_{23}& s_{23}e^{i\delta}\\
  0& -s_{23}e^{-i\delta} &  c_{23} \end{pmatrix} \text{or}~ R_{13}(\theta_{13},\delta)= \begin{pmatrix}c_{13}& 0& s_{13}e^{i\delta}\\
  0&1&0\\
  -s_{13}e^{-i\delta}& 0& c_{13} \end{pmatrix}\text{or}~
 R_{12}(\theta_{12},\delta)= \begin{pmatrix} c_{12} & s_{12}e^{i\delta} & 0\\
  -s_{12}e^{-i\delta} & c_{12} & 0\\
  0&0&1 \end{pmatrix}.\end{scriptsize}
$$

H. Fritzsch and Z. Z. Xing \cite{fx}, after an analysis of the 12 combinations given by C. Jarlskog, mentioned in equation (\ref{eq.jar}), found that only 9 out of these are `structurally' different. They also noted that the phase factor $\delta$ can be associated in 3 different manners with any of the rotation matrix, however, it can be shown that these are all equivalent due to the facility of rephasing invariance. The 9 possible representations of CKM matrix given by them are shown in Table  \ref{tab:fx}.

A. Rasin \cite{ar} had also attempted to construct  possible representations of the CKM matrix using the unitarity constraints of the CKM matrix,  the different possibilities, without changing their notations, have been presented  in Table \ref{tab:rasin}.
We have closely examined these representations and find that only 6 of these are independent. For example, the representation 9 of Table \ref{tab:rasin} can be obtained from representation 4 by re-designating angle $\theta_{23}$ of representation 4 as $\theta_{13}$, as well as by changing the quadrant of the angles  $\theta_{12}$ and $\theta_{12}^\prime$, the representation 4 thus becomes 
\be 
\begin{pmatrix}
-s_{12}s_{12}^\prime + c_{12}c_{12}^\prime c_{13} e^{-i\delta}  &  s_{12}c_{12}^\prime +c_{12}c_{13}s_{12}^\prime e^{-i\delta}  &s_{13}c_{12} e^{-i\delta} \\
 - c_{12}s_{12}^\prime e^{i\delta} - c_{13}s_{12}c_{12}^\prime  & c_{12}c_{12}^\prime e^{i\delta}  -c_{13}s_{12}s_{12}^\prime  &-s_{12}s_{13}\\
 -s_{13}c_{12}^\prime & -s_{13}s_{12}^\prime & c_{13} 
\end{pmatrix}.
\ee
Using the facility of rephasing invariance, multiplying the above matrix from the left side by the  matrix
 \be \begin{pmatrix}
e^{i\delta}&0&0\\
0& e^{-i\delta}&0\\
0&0&1
\end{pmatrix} \ee 
one obtains representation 9.
   Similarly, representations 5 and 6 are related to 7 and 8 respectively. 
\begin{table}[!]
\renewcommand{\arraystretch}{1.5}
\centering
\caption{Representations of the CKM matrix given by Ref. \cite{fx} in terms of the three mixing angles $\theta_{12}, \theta_{23}, \theta_{13}$ and CP violating phase $\delta$.}
\vspace{0.5cm}

\begin{tabular}{|c|c|c|}
\hline
S.No. & Product of rotation matrices & Resultant  Matrix\\ \hline

  1& $R_{12}(\theta_{12}) R_{23}(\theta_{23},\delta ) R_{12}^{-1}(\theta_{12}^\prime) $&$\begin{pmatrix}
c_{23} s_{12} s_{12}^\prime +c_{12} c_{{12}}^\prime e^{-i\delta}&c_{23}  s_{12} c_{12}^\prime -c_{12} s_{12}^\prime e^{-i\delta}&s_{23}  s_{12} \\

c_{23}  c_{12} s_{12}^\prime -s_{12} c_{12}^\prime e^{-i\delta}&c_{23}  c_{12} c_{12}^\prime +s_{12} s_{12}^\prime e^{-i\delta}&s_{23}  c_{12} \\

-s_{23} s_{12}^\prime &-s_{23}  c_{12}^\prime& c_{23}   \end{pmatrix} $ \\ \hline 

 2 & $ R_{23}(\theta_{23}) R_{12}(\theta_{12},\delta ) R_{23}^{-1}(\theta_{23}^\prime) $ & $ \begin{pmatrix}
  c_{12} &s_{12} c_{23} ^\prime &-s_{12} s_{23} ^\prime\\
  -s_{12} c_{23}  &c_{12} c_{23}^\prime c_{23}  +s_{23} ^\prime s_{23}  e^{-i\delta} & -c_{12} s_{23} ^\prime c_{23}  +c_{23} ^\prime s_{23}  e^{-i\delta}\\
  s_{12} s_{23}  & -c_{12} c_{23} ^\prime s_{23}  +s_{23} ^\prime c_{23}  e^{-i\delta} & c_{12} s_{23} ^\prime s_{23} +c_{23} ^\prime c_{23}  e^{-i\delta}\end{pmatrix} $ \\ \hline
  
  3& $  R_{23}(\theta_{23}) R_{13}(\theta_{13},\delta ) R_{12}(\theta_{12}) $ & $\begin{pmatrix}
 c_{13}  c_{12} &c_{13} s_{12} &s_{13} \\
 -s_{13} c_{12} s_{23}  -s_{12} c_{23}  e^{-i\delta}& -s_{13} s_{12} s_{23}  +c_{12} c_{23}  e^{-i\delta}&c_{13} s_{23}  \\
 -s_{13} c_{12} c_{23}  +s_{12} s_{23}  e^{-i\delta}&-s_{13} s_{12} c_{23}  -c_{12} s_{23}  e^{-i\delta}&c_{13} c_{23} 
\end{pmatrix} $ \\ \hline

 4& $   R_{12}(\theta_{12}) R_{13}(\theta_{13},\delta ) R_{23}^{-1}(\theta_{23}) $ & $ \begin{pmatrix}
    c_{13} c_{12} & s_{13} c_{12} s_{23}  +s_{12} c_{23}  e^{-i\delta}&s_{13} c_{12} c_{23}  -s_{12} s_{23}  e^{-i\delta}\\
    -c_{13} s_{12} &-s_{13} s_{12} s_{23}  +c_{12} c_{23}  e^{-i\delta}&-s_{13} s_{12} c_{23}  -c_{12} s_{23}  e^{-i\delta}\\
   - s_{13} &c_{13} s_{23} &c_{13} c_{23}  \end{pmatrix} $\\ \hline
 5& $ R_{13}(\theta_{13}) R_{12}(\theta_{12},\delta ) R_{13}^{-1}(\theta_{13}^\prime)$ & $\begin{pmatrix}
    c_{12}c_{13} c_{13}^\prime +s_{13} s_{13}^\prime e^{-i\delta}&s_{12} c_{13} &-c_{12} c_{13} s_{13}^\prime +s_{13}c_{13}^\prime e^{-i\delta}\\
    -s_{12} c_{13}^\prime &c_{12} &s_{12} s_{13}^\prime \\
    -c_{12} s_{13} c_{13}^\prime +c_{13} s_{13}^\prime e^{-i\delta}&-s_{12} s_{13} &c_{12}s_{13} s_{13}^\prime +c_{13} c_{13}^\prime e^{-i\delta} \end{pmatrix} $ \\ \hline
     6 & $ R_{12}(\theta_{12}) R_{23}(\theta_{23},\delta ) R_{13}(\theta_{13})$ & $\begin{pmatrix}
  -s_{23}  s_{13} s_{12} +c_{13} c_{12} e^{-i\delta}&c_{23} s_{12} &s_{23}  c_{13} s_{12} +s_{13} c_{12} e^{-i\delta}\\
  -s_{23}  s_{13} c_{12} -c_{13} s_{12} e^{-i\delta}&c_{23}  c_{12} &s_{23}  c_{13} c_{12} -s_{13} s_{12} e^{-i\delta}\\
  -c_{23}  s_{13} &-s_{23}  &c_{23}  c_{13}  \end{pmatrix} $ \\ \hline
    7& $ R_{23}(\theta_{23}) R_{12}(\theta_{12},\delta ) R_{13}^{-1}(\theta_{13})$ & $\begin{pmatrix} 
c_{12} c_{13} &s_{12}&-c_{12} s_{13} \\
-s_{12} c_{13} c_{23}  +s_{13} s_{23}  e^{-i\delta}&c_{12} c_{23}  &s_{12} s_{13} c_{23}  +c_{13}s_{23}  e^{-i\delta}\\
s_{12}c_{13} s_{23}  +s_{13} c_{23}  e^{-i\delta}&-c_{12} s_{23}  &-s_{12} s_{13} s_{23}  +c_{13} c_{23}  e^{-i\delta}
\end{pmatrix}  $ \\ \hline
    8& $ R_{13}(\theta_{13}) R_{12}(\theta_{12},\delta ) R_{23}(\theta_{23})$ & $
 \begin{pmatrix}
c_{12} c_{13} &s_{12} c_{13} c_{23}  -s_{13} s_{23}  e^{-i\delta}&s_{12} c_{13} s_{23}  +s_{13} c_{23}  e^{-i\delta}\\
-s_{12} &c_{12} c_{23}  &c_{12} s_{23} \\
-c_{12} s_{13} &-s_{12} s_{13} c_{23}  -c_{13} s_{23}  e^{-i\delta}&
 -s_{12} s_{13} s_{23}  +c_{13} c_{23}  e^{-i\delta}
\end{pmatrix} $ \\ \hline
9& $R_{13}(\theta_{13}) R_{23}(\theta_{23},\delta ) R_{12}^{-1}(\theta_{12}) $ & $\begin{pmatrix}
-s_{23}  s_{13} s_{12}+c_{13} c_{12} e^{-i\delta}  &-s_{23}  s_{13} c_{12} -c_{13} s_{12} e^{-i\delta}&c_{23}  s_{13}\\
c_{23}  s_{12} & c_{13} c_{12} & s_{23}  \\
-s_{23}  c_{13}s_{12} -s_{13} c_{12} e^{-i\delta}&-s_{23}  c_{13} c_{12} +s_{13} s_{12} e^{-i\delta}&c_{23}  c_{13} \end{pmatrix} $\\ \hline

\end{tabular} 
\label{tab:fx} 

\end{table}

 \begin{table}[!htp]
 \renewcommand{\arraystretch}{1.2}

\caption{Representations of the CKM matrix given by Ref. \cite{ar} in terms of the three mixing angles $\theta_{12}, \theta_{23}, \theta_{13}$ and CP violating phase $\delta$.}
\vspace{0.5cm}
\centering
\begin{tabular}{|c|c|}
\hline
S.No. & Resultant  Matrices\\ \hline
1&$ \begin{pmatrix}
 c_{12}c_{13}&c_{13}s_{12}&s_{13}e^{-i\delta}\\
 -s_{13}c_{12}s_{23}e^{i\delta}-s_{12}c_{23}& -s_{12}s_{23}s_{13}e^{i\delta}+c_{12}c_{23}&c_{13}s_{23}\\
 -s_{13}c_{12}c_{23}e^{i\delta}+s_{12}s_{23}&-s_{12}s_{13}c_{23}e^{i\delta}-c_{12}s_{23}&c_{23}c_{13}
 
\end{pmatrix} $ \\ \hline

 2& $\begin{pmatrix} 
c_{12}c_{13}&s_{12}e^{-i\delta}&c_{12}s_{13}\\
-s_{12}c_{23}c_{13}e^{i\delta}-s_{13}s_{23}&c_{12}c_{23}&-s_{13}s_{12}c_{23}e^{i\delta}+c_{13}s_{23}\\
s_{12}c_{13}s_{23}e^{i\delta}-s_{13}c_{23}&-c_{12}s_{23}&s_{13}s_{23}s_{12}e^{i\delta}+c_{23}c_{13}

\end{pmatrix}  $ \\ \hline
 
3& $ \begin{pmatrix}
s_{12}s_{23}s_{13}e^{-i\delta}+c_{12}c_{23}  &-s_{23}s_{13}c_{12}e^{-i\delta}+c_{13}s_{12}&c_{23}s_{13}e^{-i\delta}\\
-c_{23}s_{12}& c_{12}c_{23}& s_{23}\\
s_{12}s_{23}c_{13}-s_{13}c_{12}e^{i\delta}&-s_{23}c_{12}c_{13}-s_{12}s_{13}e^{i\delta}&c_{13}c_{23} 
\end{pmatrix} $\\ \hline

4&$\begin{pmatrix}
-c_{23}s_{12}s_{12}^\prime e^{-i\delta}+c_{12}c_{12}^\prime &c_{23}s_{12}c_{12}^\prime e^{-i\delta}+c_{12}s_{12}^\prime &s_{12}s_{23}e^{-i\delta}\\
-c_{12}c_{23}s_{12}^\prime-s_{12}c_{12}^\prime e^{i\delta}&c_{12}c_{23}c_{12}^\prime-s_{12}s_{12}^\prime e^{i\delta}&s_{23}c_{12}\\
s_{12}^\prime s_{23}&-s_{23}c_{12}^\prime &c_{23} 
\end{pmatrix} $ \\ \hline 

5& $\begin{pmatrix}
    -c_{23}s_{13}s_{13}^\prime e^{-i\delta}+c_{13}c_{13}^\prime &-s_{13}s_{23}e^{-i\delta}&c_{13}^\prime s_{13}c_{23}e^{-i\delta}+c_{13}s_{13}^\prime \\
      -s_{13}^\prime s_{23}&c_{23}&s_{23}c_{13}^\prime\\
 -c_{13}c_{23}s_{13}^\prime-s_{13}c_{13}^\prime e^{i\delta}&-s_{23}c_{13}& c_{13}c_{23}c_{13}^\prime - s_{13}s_{13}^\prime e^{i\delta}
\end{pmatrix}$ 
 \\ \hline
6 & $\begin{pmatrix}
  c_{12}&s_{12}c_{23}^\prime &s_{12}s_{23}^\prime\\
 - s_{12}c_{23}&c_{12}c_{23}c_{23}^\prime-s_{23}s_{23}^\prime e^{i\delta}&c_{12}s_{23}^\prime c_{23}+c_{23}^\prime s_{23}e^{i\delta}\\
  s_{12}s_{23}e^{-i\delta} & -c_{12}c_{23}^\prime s_{23}e^{-i\delta}- s_{23}^\prime c_{23}&-c_{12}s_{23}s_{23}^\prime e^{-i\delta}+c_{23}c_{23}^\prime
\end{pmatrix} $ \\ \hline

7&  $\begin{pmatrix}
    c_{13}c_{12}c_{13}^\prime - s_{13}s_{13}^\prime e^{i\delta}& s_{12}c_{13} & c_{13}c_{12}s_{13}^\prime + s_{13}c_{13}^\prime e^{i\delta}\\
    -s_{12}c_{13}^\prime & c_{12} & -s_{12}s_{13}^\prime\\
    -c_{12}s_{13}c_{13}^\prime e^{-i\delta} - c_{13}s_{13}^\prime & -s_{13}s_{12}e^{-i\delta} & -c_{12}s_{13}s_{13}^\prime e^{-i\delta}+c_{13}c_{13}^\prime  
\end{pmatrix} $ \\ \hline

8& $\begin{pmatrix}
  c_{13}& -s_{13}s_{23}^\prime & s_{13}c_{23}^\prime\\
 - s_{13}s_{23} e^{-i\delta} &-c_{13}s_{23}s_{23}^\prime e^{i\delta} +c_{23}c_{23}^\prime &c_{13}c_{23}^\prime s_{23} e^{i\delta} + s_{23}^\prime c_{23}\\
-  s_{13}c_{23} &-c_{13}s_{23}^\prime c_{23}-c_{23}^\prime s_{23} e^{-i\delta}& c_{13}c_{23}c_{23}^\prime - s_{23}s_{23}^\prime e^{-i\delta}
\end{pmatrix} $\\ \hline
  
9& $ \begin{pmatrix}
 c_{12}c_{12}^\prime c_{13}-s_{12}s_{12}^\prime e^{i\delta}& c_{12}c_{13}s_{12}^\prime + s_{12}c_{12}^\prime e^{i\delta}&s_{13}c_{12}\\
 -c_{13}s_{12}c_{12}^\prime e^{-i\delta} - c_{12}s_{12}^\prime & -c_{13}s_{12}s_{12}^\prime e^{-i\delta} + c_{12}c_{12}^\prime &-s_{12}s_{13}e^{-i\delta}\\
 -s_{13}c_{12}^\prime & -s_{13}s_{12}^\prime & c_{13} 
\end{pmatrix}$\\ \hline
 \end{tabular}
  \label{tab:rasin}
  \end{table}
   \clearpage
   
In an another approach \cite{abg, beta}, attempt has been made to use experimentally measurable quantities, i.e., magnitudes of the CKM matrix elements and angles  $\alpha$, $\beta$ or $\gamma$ of the unitarity triangle as the CP violating phase  of the CKM matrix  resulting into several possible representations of the CKM matrix involving 4 measurable parameters. Again to facilitate discussion, we reproduce some essentials of these attempts here. 
For example, considering
angle $\gamma$ as the phase of the CKM matrix, 4 parametrizations have been obtained, referred to as the $\gamma $ angle parametrizations.
This angle can be expressed in terms of the elements of the CKM matrix as
   \be \gamma= arg\left(-\frac{V_{ud} V_{ub}^*}{V_{cd}V_{cb}^*}\right). \label{g}\ee
   The phase $\gamma$ can be allocated along with either of the 4 CKM matrix elements appearing in the definition of $\gamma$, i.e., $V_{ud,ub,cd,cb}$, leading to only one of these being complex and all others being real and positive, e.g., 
   \begin{align*}
  & \gamma_1~:~(|V_{ud}|,~|V_{ub}|,~|V_{cd}|,~-|V_{cb}|e^{i\gamma}),\\
  & \gamma_2~:~(|V_{ud}|,~|V_{ub}|,~-|V_{cd}|e^{-i\gamma},~|V_{cb}|),\\
  & \gamma_3~:~(|V_{ud}|,~-|V_{ub}|e^{-i\gamma},~|V_{cd}|,~|V_{cb}|),\\
   & \gamma_4~:~(-|V_{ud}|e^{i\gamma},~|V_{ub}|,~|V_{cd}|,~|V_{cb}|).
   \end{align*}
   The above defines 4  parametrizations($\gamma_1$, $\gamma_2$, $\gamma_3$ and $\gamma_4$) of the CKM matrix in which $\gamma$ is explicitly the CP violating phase, all 
4 of these   being equivalent. To obtain parametrization $\gamma_3$,
one can use $\gamma$, $|V_{cd}|$, $|V_{cs}|$, $|V_{td}|$  as independent variables and express others as functions of them, resulting into
   \be
  V^{\gamma_3}_{CKM}=\begin{pmatrix}
  |V_{ud}|&-\frac{|V_{ud}||V_{cd}|-|V_{ub}||V_{cb}|e^{-i\gamma}}{|V_{cs}|}& -|V_{ub}|e^{-i\gamma}\\
   |V_{cd}|&|V_{cs}|& |V_{cb}|\\
    |V_{td}|&   \frac{(|V_{cb}|^2-|V_{td}|^2)|V_{cd}|-|V_{cb}||V_{ud}||V_{ub}|e^{-i\gamma}}{|V_{cs}||V_{td}|}&\frac{|V_{ud}||V_{ub}|e^{-i\gamma}-|V_{cd}||V_{cb}|}{|V_{td}|}\\
  \end{pmatrix},\label{gamma}
   \ee where $ |V_{ud}|=\sqrt{1-|V_{cd}|^2-|V_{td}|^2}$, ~~ $ |V_{cb}|=\sqrt{1-|V_{cd}|^2-|V_{cs}|^2}$, ~~
   \vspace{0.2cm}\\
 $   ~~~~~~~~~~ |V_{ub}|=\frac{|V_{ud}||V_{cd}||V_{cb}|\cos \gamma}{1-|V_{cd}|^2}- \sqrt{\left(\frac{ |V_{ud}||V_{cd}||V_{cb}|\cos\gamma }{1-|V_{cd}|^2}\right)^2-\frac{|V_{cs}|^2(|V_{ud}|^2-1)+|V_{ud}|^2|V_{cd}|^2}{1-|V_{cd}|^2}}
   $. \vspace{0.2cm}\\ 
 The other 3 $\gamma$ parametrizations can be obtained in a similar manner. Considering angles $\alpha$ and $\beta$ as the CP violating phase of the CKM matrix, the corresponding $\alpha $ and $\beta$ parametrizations can also be  obtained.
 Interestingly, the authors have shown that the 4 parametrizations for $\alpha$ or $\beta$ or $\gamma$ are all equivalent and also these 12 parametrizations can be transformed from each other and again are all equivalent.

   \section{Cartesian derivation of independent representations of the CKM matrix}
To understand the issue of construction of 9 independent representations of the $V_{CKM}$, we have attempted to carry out this task in a rigorous ab-initio manner, without involving the rotation matrices, henceforth these would be referred to as Cartesian representations. To this end, we  follow an approach wherein the 9 independent representations are constructed using any individual element of a $3\times3$ complex unitary matrix V given by
\begin{equation}
V= \begin{pmatrix}
a_{11} & a_{12} & a_{13}\\
a_{21} & a_{22} & a_{23}\\
a_{31} & a_{32} & a_{33}
\end{pmatrix}.\label{eq.gmatrix}
\end{equation}
 It may be noted that the CKM matrix is sandwiched between quark fields which allows 5 out of 6 phases of above $3\times3$ unitary matrix to be removed using rephasing invariance, leaving the  matrix having 3 independent angles and 1 non removable phase. Further, the elements of the CKM matrix should obey the  following unitarity constraints
\begin{equation}
 \displaystyle{\sum \limits_{i =1}^3}a_{\alpha i}a_{\beta i}^* = \delta_{\alpha\beta},
 \hspace{1.5cm} \displaystyle{\sum \limits_{\alpha=1}^3}a_{\alpha i}a_{\alpha j}^* = \delta_{ij},\label{eq.uni}
  \end{equation}
  where $\alpha, ~\beta\equiv(1,~2,~3)$ and $i,~j\equiv (1,~2,~3)$. Taking into consideration the physical structure of CKM matrix, while constructing its representations one needs to consider the diagonal elements of matrix V, given in equation (\ref{eq.gmatrix}), to be nearly equal to unity whereas the off diagonal elements should be much smaller than unity.
  
To illustrate our procedure, we consider an example wherein we begin with a complex element $a_{21}$ of the matrix V, defined as
 \begin{equation}
a_{21}\equiv s_1e^{i\phi_{21}}, ~~~\text{where}~ s_1=\sin\theta_1. \label{eq.start}
\end{equation}
Following the unitarity constraints, one may introduce two more angles $\theta_2$ and $\theta_3$ such that
\begin{equation}
a_{11}=c_1c_2 e^{i\phi_{11}}, \hspace{0.5cm}  a_{31}=c_1s_2 e^{i\phi_{31}}, \hspace{0.5cm} a_{22}=c_1c_3 e^{i\phi_{22}},\hspace{0.5cm} a_{23}=c_1s_3 e^{i\phi_{23}}, \label{eq.4}
\end{equation}
where $c_i=\cos\theta_i$ and $s_i=\sin\theta_i$, with i = 1, 2, 3.
It is interesting to note that this is a unique way to define the above  elements in terms of the mixing angles and any other way would disturb the unitarity relations. Using these, the matrix V given in equation (\ref{eq.gmatrix}) becomes
\begin{equation}
V=\begin{pmatrix}
c_1c_2 e^{i\phi_{11}} & a_{12} & a_{13}\\
s_1 e^{i\phi_{21}} &c_1c_3 e^{i\phi_{22}} & c_1s_3 e^{i\phi_{23}}\\
c_1s_2 e^{i\phi_{31}} & a_{32} & a_{33}.
\end{pmatrix}\end{equation}
The above equation can also be written as
\begin{equation} 
V=\begin{pmatrix}
e^{i\phi_{11}} &0 & 0\\
0 & e^{i\phi_{21}} & 0\\
0 &0 & e^{i\phi_{31}}
\end{pmatrix}
\begin{pmatrix}
c_1c_2  & a_{12}e^{i(\phi_{21}-\phi_{22}-\phi_{11})} & a_{13}e^{i(\phi_{21}-\phi_{23}-\phi_{11})}\\
s_1  &c_1c_3  & c_1s_3\\
c_1s_2  & a_{32}e^{i(\phi_{21}-\phi_{22}-\phi_{31})} & a_{33}e^{i(\phi_{21}-\phi_{23}-\phi_{31})}
\end{pmatrix}
\begin{pmatrix}
1 &0 &0\\
0  & e^{i(\phi_{22}-\phi_{21})} & 0\\
0 & 0 & e^{i(\phi_{23}-\phi_{21})}
\end{pmatrix}.
\end{equation}
Further, using unitarity conditions mentioned in equation (\ref{eq.uni}), one gets 
   \begin{equation} 
V=\begin{pmatrix}
e^{i\phi_{11}} &0 & 0\\
0 & e^{i\phi_{21}} & 0\\
0 &0 & e^{i\phi_{31}}
\end{pmatrix}
\begin{pmatrix}
c_1c_2  & a_{12}& a_{13}\\
s_1  &c_1c_3  & c_1s_3\\
c_1s_2  & a_{32} & a_{33}
\end{pmatrix}
\begin{pmatrix}
1 &0 &0\\
0  & e^{i(\phi_{22}-\phi_{21})} & 0\\
0 & 0 & e^{i(\phi_{23}-\phi_{21})}
\end{pmatrix}. \label{ps}
\end{equation}

As a next step, we now derive the remaining 4 complex elements of the CKM matrix, i.e., $a_{12}$, $a_{13}$, $a_{32}$ and $a_{33}$. Using the already defined 5 elements of the CKM matrix, mentioned in equations (\ref{eq.start}) and (\ref{eq.4}), the following unitarity constraints
 \be \begin{array}{l} a_{11}a_{21}^*+a_{12}a_{22}^*+a_{13}a_{23}^*=0,\\
 |a_{11}|^2 +|a_{12}|^2 +|a_{13}|^2=1, \end{array}\ee
can be re-written as
  \begin{equation}\begin{array}{l} 
c_1c_2s_1+a_{12}c_1c_3+a_{13}c_1s_3=0,\\ c_1^2c_2^2+|a_{12}|^2 +|a_{13}|^2 =1.\end{array}
  \end{equation}
 By splitting the complex elements $a_{12}$, $a_{13}$ into real and imaginary parts, 
 we get
  \begin{equation}
   c_1c_2s_1+Re(a_{12})c_1c_3+ Re(a_{13})c_1s_3=0 \label{re},
  \end{equation}
    \begin{equation}
    Im(a_{12})c_1c_3+ Im(a_{13})c_1s_3=0 \label{im},
  \end{equation}
    \begin{equation}
    (Re (a_{12}))^2 + (Im (a_{12}))^2+ (Re (a_{13}))^2+ (Im (a_{12}))^2 =1-c_1^2c_2^2 \label{c}.
  \end{equation}
  From equation (\ref{im}), for non-zero $c_1$, $Im(a_{12})$ and $Im(a_{13})$ can be expressed with a real coefficient $\alpha$ as
  \be Im(a_{12})=-\alpha s_3  ~~~~\text{and} ~~~~Im(a_{13})=\alpha c_3 .\ee
 Similarly, from equation (\ref{re}),  $Re(a_{12})$ and $Re(a_{13})$ can be expressed with real coefficient $\beta$ as
   $$ Re(a_{12})=-\beta s_3 - s_1c_2c_3 ~~~~\text{and} ~~~~Re(a_{13})=\beta c_3 - s_1c_2s_3  .$$
   Using these relations, equation (\ref{c}) becomes
   \be \alpha^2+\beta^2=1-c_1^2c_2^2-s_1^2c_2^2 = s_2^2 \ee
     \text{implying}\be
   \beta+i\alpha=-s_2 e^{-i\delta},\ee
   Therefore, the elements $a_{12}$  and $a_{13}$ can be written as
   \begin{equation}
   a_{12}= -s_1c_2c_3+s_2s_3 e^{-i\delta}~~~~\text{and} ~~~~ a_{13}=-s_1c_2s_3 -s_2c_3 e^{-i\delta}.\label{12}
   \end{equation}
   Similarly, using the unitarity relations
    \be \begin{array}{l} a_{21}a_{31}^*+a_{22}a_{32}^*+a_{23}a_{33}^*=0,\\
 |a_{31}|^2 +|a_{32}|^2 +|a_{33}|^2=1, \end{array}\ee
one obtains
  \begin{equation}
   a_{32}= -s_1s_2c_3-c_2s_3e^{-i\delta}~~~~\text{and} ~~~~ a_{33}= -s_1s_2s_3+c_2c_3e^{-i\delta}.\label{32}
   \end{equation}
Finally, using equations (\ref{12}) and(\ref{32}), matrix V can be written as 
   \begin{equation}
   \begin{pmatrix} e^{i\phi_{11}}&0&0\\0&e^{i\phi_{21}}&0\\0&0&e^{i\phi_{31}}
\end{pmatrix}
\begin{pmatrix}
c_1c_2&-s_1c_2c_3+s_2s_3e^{-i\delta}&-s_1c_2s_3-s_2c_3e^{-i\delta}\\
s_1&c_1c_3&c_1s_3\\
c_1s_2&-s_1s_2c_3-c_2s_3e^{-i\delta}&
 -s_1s_2s_3+c_2c_3e^{-i\delta}
 
\end{pmatrix}
\begin{pmatrix} 1&0&0\\0&e^{i(\phi_{22}-\phi_{21})}&0\\0&0&e^{i(\phi_{23}-\phi{21})}
\end{pmatrix}\end{equation}
The factored out phases can be removed by using the facility of rephasing invariance,  yielding the following representation of the CKM matrix in terms of 3 mixing angles and 1 non removable phase
\be 
\begin{pmatrix}
c_1c_2&-s_1c_2c_3+s_2s_3e^{-i\delta}&-s_1c_2s_3-s_2c_3e^{-i\delta}\\
s_1&c_1c_3&c_1s_3\\
c_1s_2&-s_1s_2c_3-c_2s_3e^{-i\delta}&
 -s_1s_2s_3+c_2c_3e^{-i\delta}
 
\end{pmatrix}.
\ee

Similarly, the other independent representations can also be constructed by using a different starting element of the  matrix V.
In Table \ref{tab:all}, we have summarized the 9 independent Cartesian parametrizations of the CKM matrix along with the corresponding starting element of unitary matrix V.
\begin{table}
 \renewcommand{\arraystretch}{1.2}
\centering
\caption{ Cartesian representations of the CKM matrix.}
\vspace{0.5cm}
\begin{tabular}{|c|c|c|}
\hline
S.No. & Starting  element of V & Resultant  Matrix\\ \hline
1& $ a_{13}=s_1e^{i\phi_{13}} $ & $\begin{pmatrix}
 c_1c_2&c_1s_2&s_1\\
 -s_1c_2s_3-s_2c_3e^{-i\delta}& -s_1s_2s_3+c_2c_3e^{-i\delta}&c_1s_3\\
 -s_1c_2c_3+s_2s_3e^{-i\delta}&-s_1s_2c_3-c_2s_3e^{-i\delta}&c_1c_3
\end{pmatrix} $ \\ \hline 

2&$a_{12}=s_1e^{i\phi_{12}}$ & $\begin{pmatrix} 
c_1c_2&s_1&c_1s_2\\
-s_1c_2c_3+s_2s_3e^{-i\delta}&c_1c_3&-s_1s_2c_3-c_2s_3e^{-i\delta}\\
-s_1c_2s_3-s_2c_3e^{-i\delta}&c_1s_3&-s_1s_2s_3+c_2c_3e^{-i\delta}
\end{pmatrix}  $ \\ \hline

3& $ a_{21}=s_1e^{i\phi_{21}}$ & $
 \begin{pmatrix}
c_1c_2&-s_1c_2c_3+s_2s_3e^{-i\delta}&-s_1c_2s_3-s_2c_3e^{-i\delta}\\
s_1&c_1c_3&c_1s_3\\
c_1s_2&-s_1s_2c_3-c_2s_3e^{-i\delta}&
 -s_1s_2s_3+c_2c_3e^{-i\delta}
\end{pmatrix} $ \\ \hline
4& $a_{23}=s_1e^{i\phi_{23}} $ & $\begin{pmatrix}
-s_1s_2s_3+c_2c_3e^{-i\delta}  &-s_1s_2c_3-c_2s_3e^{-i\delta}&c_1s_2\\
c_1s_3& c_1c_3& s_1\\
-s_1c_2s_3-s_2c_3e^{-i\delta}&-s_1c_2c_3+s_2s_3e^{-i\delta}&c_1c_2 \end{pmatrix} $\\ \hline
  
  5& $ a_{31}=s_1e^{i\phi_{31}} $ & $ \begin{pmatrix}
    c_1c_2& -s_1c_2s_3-s_2c_3e^{-i\delta}&-s_1c_2c_3+s_2s_3e^{-i\delta}\\
    c_1s_2&-s_1s_2s_3+c_2c_3e^{-i\delta}&-s_1s_2c_3-c_2s_3e^{-i\delta}\\
    s_1&c_1s_3&c_1c_3\end{pmatrix} $\\ \hline
   6 & $ a_{32}=s_1e^{i\phi_{32}} $ & $\begin{pmatrix}
  -s_1s_2s_3+c_2c_3e^{-i\delta}&c_1s_3&-s_1c_2s_3-s_2c_3e^{-i\delta}\\
  -s_1s_2c_3-c_2s_3e^{-i\delta}&c_1c_3&-s_1c_2c_3+s_2s_3e^{-i\delta}\\
  c_1s_2&s_1&c_1c_2 \end{pmatrix} $ \\ \hline
  
 7 & $ a_{11}=c_1 e^{i\phi_{11}} $ & $ \begin{pmatrix}
  c_1&s_1c_2&s_1s_2\\
  -s_1c_3&c_1c_2c_3-s_2s_3 e^{-i\delta} & c_1s_2c_3+c_2s_3e^{-i\delta}\\
  s_1s_3& -c_1c_2s_3-s_2c_3e^{-i\delta} & -c_1s_2s_3+c_2c_3e^{-i\delta}\end{pmatrix} $ \\ \hline

   8& $ a_{22}=c_1e^{i\phi_{22}} $ & $\begin{pmatrix}
    c_1c_2c_3-s_2s_3e^{-i\delta}&s_1c_2&-c_1c_2s_3-s_2c_3e^{-i\delta}\\
    -s_1c_3&c_1&s_1s_3\\
    c_1s_2c_3+c_2s_3e^{-i\delta}&s_1s_2&-c_1s_2s_3+c_2c_3e^{-i\delta} \end{pmatrix} $ \\ \hline
   
  9& $a_{33}=c_1e^{i\phi_{33}} $ &$\begin{pmatrix}
-c_1s_2s_3+c_2c_3e^{-i\delta}&c_1s_2c_3+c_2s_3e^{-i\delta}&s_1s_2\\
-c_1c_2s_3-s_2c_3e^{-i\delta}&c_1c_2c_3-s_2s_3e^{-i\delta}&s_1c_2\\
s_1s_3&-s_1c_3&c_1 \end{pmatrix} $ \\ \hline 
\end{tabular} 
\label{tab:all}  
\end{table}

As a next step, it is desirable to explore their relationship with the representations available in literature.  For example, considering the representation 1 of the CKM matrix given in Table \ref{tab:all}, one can obtain the commonly used parametrization \cite{chau} adopted by PDG \cite{pdg22},  e.g., 
\begin{equation}
V_{CKM} = \bp
c_{12}c_{13} & s_{12}c_{13} & s_{13}e^{-i\delta} \\ 
-s_{12}c_{23}-c_{12}s_{23}s_{13}e^{i\delta} & c_{12}c_{23}-s_{12}s_{23}s_{13}e^{i\delta} & s_{23}c_{13} \\ 
s_{12}s_{23}-c_{12}c_{23}s_{13}e^{i\delta} & -c_{12}s_{23}-s_{12}c_{23}s_{13}e^{i\delta} & c_{23}c_{13}
\ep , \label{eq.pdg}
\end{equation}
where $c_{ij}= \cos \theta_{ij}, ~ s_{ij}=\sin \theta_{ij}$ for i, j=1, 2, 3, with $\theta_{12}, ~\theta_{23}$, ~$\theta_{13}$ and $\delta$ being the 3 mixing angles  and  the CP violating phase respectively.
To do so, we need to re-designate $s_1\rightarrow s_{13}$,  $s_2\rightarrow s_{12}$ and $s_3\rightarrow s_{23}$ as well as carry out rephasing of the quark fields using the multiplication of matrices
$$\begin{pmatrix}
e^{-i\delta}&0&0\\
0&1&0\\
0&0&1
\end{pmatrix} ~~~~\text{ and}~~~~~~ \begin{pmatrix}
e^{i\delta}&0&0\\
0&e^{i\delta}&0\\
0&0&1
\end{pmatrix},
$$
respectively on the left and right side of Cartesian representation 1. The Kobayashi-Maskawa (KM) representation \cite{ckm2,kmrep} can be shown to be related to the Cartesian representation 7.

A look at the representations constructed here, given in Table \ref{tab:all}, and the ones given by Ref.~\cite{fx} in Table \ref{tab:fx} reveal that our representations 1, 2, 3, 4, 5, 6, 7, 8 and 9 are related to 3, 7, 8, 9, 4, 6, 2, 5 and 1 respectively. Similarly, on comparison with representations given by Ref.~\cite{ar} in Table \ref{tab:rasin}, it is  found that our parametrizations 1, 2, 4, 7, 8 and 9   are related to  1, 2, 3, 6, 7 and 4 of Table \ref{tab:rasin} respectively. It needs to be emphasized that there are no representations in Table \ref{tab:rasin} which corresponds to Cartesian representations 3, 5 and 6.

\clearpage

\section{Unitarity based numerical evaluation of the Cartesian representations } 
\begin{figure}
 \centering
  \includegraphics[scale=0.65]{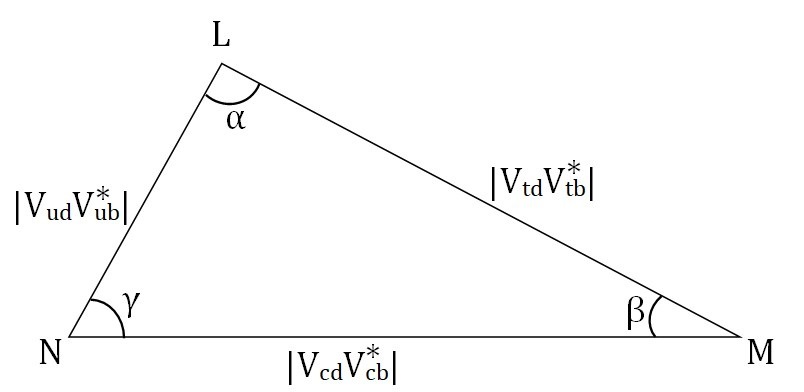}  
  \caption{ The db unitarity triangle.}
  \label{db}
 \end{figure}

After having discussed the relationship of 9 Cartesian representations with other similar representations \cite{jar}-\cite{beta}, we have carried out a unitarity based analysis, using the latest data, to understand the significance of these representations.
 To begin with, we have calculated mixing angles and CP violating phase for each parametrization for numerical evaluation of the corresponding CKM matrix. 
As a first step, we have considered the representation 1  of Table 3, this being equivalent  to the PDG representation. In order to find $\theta_1$, we have evaluated $|V_{ub}|$ using a unitarity based analysis involving  the `db' triangle \cite{pap}, shown in Figure \ref{db}.
 From this triangle, using the relationship between its angles and sides, one gets
  \be
|V_{ub}|= \frac{|V_{cd}||V_{cb}|\sin\beta}{|V_{ud}|\sin\alpha}.\ee
Using the PDG values \cite{pdg22} of $|V_{cd}| = 0.221 \pm 0.004$, $|V_{ud}|=0.97373 \pm 0.00031$ and the value of $|V_{cb}|=(40.6\pm 0.9)\times 10^{-3}$ as advocated by Belle collaboration \cite{belle23}
as well as the values of $\alpha$ and $\beta$  as given by PDG  \cite{pdg22}, 
 we get \be s_1 = |V_{ub}|=(3.4939\pm 0.1465)\times 10^{-3},\label{eq.ub}~\text{implying}~ \theta_1=   0.00349 \pm 0.00015. \ee
This is a rigorous unitarity based value of $V_{ub}$, which is in agreement with values given in Refs.~\cite{pap,nik}. This value of $V_{ub}$ implies the ratio $ \left|\frac{V_{ub}}{V_{cb}}\right|=0.08506\pm0.00373$,  in agreement with  measurements from $\Lambda_b\rightarrow\rho \mu\nu$ and $B_s\rightarrow K\mu \nu$ decays \cite{hflav23}. 
Further, considering the latest PDG value \cite{pdg22} of $|V_{us}|$, one gets 
\be
 s_{2} = \frac{|V_{us}|}{c_1}=\frac{0.2243\pm 0.0008}{c_1},
 ~\text{implying}~
 \theta_2 = 0.2262 \pm 0.0008. \ee
 Considering the value of $|V_{cb}|$ as mentioned above, we obtain
\be
 s_{3}=\frac{|V_{cb}|}{c_1}= \frac{(40.6\pm 0.9)\times 10^{-3}}{c_1}, ~ \text{implying}~
 \theta_3 = 0.0406 \pm 0.0009. \ee

As a next step, to evaluate the phase $\delta$ corresponding to this representation, we first discuss the relationship of phase $\delta$ with the angle $\gamma$ of the unitarity triangle. To do so, one can express $\gamma$  in terms of  the elements of the mixing matrix, mentioned in equation (\ref{g}).
 This can be further expressed as
\begin{align}
\gamma  &= \arg\left[- \frac{c_{2} c_{1}(s_{1} e^{+i\delta})}{(-s_{2} c_{3}-c_{2} s_{3} s_{1}e^{i\delta})s_{3} c_{1}}\right]\\
&= \arg\left[ \frac{c_{2} (s_{2} c_{3} s_{1} \cos{\delta} + c_{2} s_{3} s^2 _{1} + i(s_{2} c_{3} s_{1} \sin{\delta} )}{s_{3}((s_{2} c_{3}+c_{2} s_{3} s_{1} \cos{\delta} )^2+( c_{2} s_{3} s_{1} \sin \delta)^2) }\right]\\
& = \tan ^{-1} \left[ \frac{s_{2} c_{3} \sin{\delta}}{s_{2} c_{3} \cos{\delta} + c_{2} s_{3} s _{1} } \right].\label{eq:gamma}
 \end{align}
 The above equation can be simplified as
\be
s_{2} c_{3} \cos{\delta} \sin \gamma + c_{2} s_{3} s _{1}\sin \gamma = s_{2} c_{3} \sin{\delta} \cos \gamma\ee which leads to the relation
\be 
\frac{\sin (\delta- \gamma)}{\sin \gamma } = \frac{c_{2} s_{3} s _{1} }{s_{2} c_{3} }.
\ee
This can be solved further to obtain 
\be 
\delta= \gamma + \sin ^{-1}\left(\sin \gamma \frac{s_1c_2s_3}{s_2c_3}\right).\label{delta}
\ee
 For the present unitarity based analysis, the angle  $\gamma$ can be found using the closure property of the angles of the unitarity triangle yielding $\gamma = (72.6 \pm 4.6)\degree$. Using the above expression one gets
\be
\delta = (72.6 \pm 4.6)\degree, 
  \ee
  implying that for representation 1 of Table 3, the CP violating phase $\delta$ be considered to be equal to the angle $\gamma$ of the unitarity triangle.

 After having found the three mixing angles and the phase $\delta$, we obtain the corresponding CKM matrix for the representation, i.e.,
\be V_{CKM}=\begin{pmatrix}
0.97451\pm0.00018 &0.2243\pm 0.0008 &0.00349\pm 0.00015\\
0.2242\pm0.0008&0.97371\pm0.00019&0.0406\pm0.0009\\
0.00872\pm0.00020&0.03981\pm0.00088& 0.99917\pm 0.00004
\end{pmatrix}\label{eq.ckm1}
 \ee 
 
A look at this matrix reveals that this shows an excellent overlap with the one obtained by PDG\cite{pdg22}
\be 
\begin{pmatrix}
0.97435\pm0.00016 &0.22500\pm 0.00067 &0.00369\pm 0.00011\\
0.22486\pm0.00067&0.97349\pm0.00016&0.04182^{+0.00085}_{-0.00074}\\
0.00857^{+0.00020}_{-0.00018}&0.04110^{+0.00083}_{-0.00072}& 0.999118^{+0.000031}_{-0.000036}
\end{pmatrix}. \ee
It needs to be emphasized that the matrix given in equation (\ref{eq.ckm1}) has been obtained using well measured CKM parameters and the unitarity based constraints. It is interesting to mention that in the representation considered by us, the hierarchy  of the CKM matrix elements is very well captured by the hierarchy of the mixing angles, i.e., $|V_{us}| \sim s_{2} \gg |V_{cb}| \sim s_{3} \gg |V_{ub}| \sim s_{1}$ . This, as well as unitarity, leads to hierarchy amongst the 9 elements of the CKM matrix, as defined in Refs. \cite{asy1}-\cite{asy3}.

 After numerically constructing this representation of the CKM matrix, as a next step we have found the angles and phases of the remaining  parametrizations in order to arrive at the numerical values of the corresponding matrix elements. The numerical evaluation of other representations is not straight forward as in these cases the CP violating phase cannot be considered to be nearly equal to the angle $\gamma$ of the unitarity triangle. Therefore, for these representations, along with  $|V_{us}|$, $|V_{cb}|$ and $|V_{ub}|$ as inputs, instead of phase $\delta$, we consider the numerical value of the element $|V_{td}|$ from the CKM matrix given in equation (\ref{eq.ckm1}). It may be noted that the element   $|V_{td}|$ captures the effects of CP violating phase $\delta$ as is emphasized in the literature \cite{review}. Using these inputs, for each Cartesian parametrization, we can then find the values of the 3 mixing angles, $\theta_1$, $\theta_2$, $\theta_3$ and the CP violating  phase $\delta$, these have been presented in column 2 of Table \ref{analysis}. It may be noted that for different parametrizations, magnitudes of the corresponding CKM matrix elements have not been given here as these are rephasing invariant quantities. 
 \begin{table}[!]
   \renewcommand{\arraystretch}{1.8} 
\setlength{\tabcolsep}{2pt}
\centering
\caption{Calculated parameters using different representations}
\vspace{0.5cm}
 \renewcommand{\arraystretch}{2.8}
\begin{tabular}{|c|c|c|}
\hline
Representation & Mixing angles and phase &Jarlskog's Invariant `J'\\ \hline
1 & \makecell{~~~$\theta_1=0.00349\pm0.00015$ \\$\theta_2=0.2262\pm0.0008$\\$\theta_3=0.0406\pm0.0009$\\ $\delta=(72.6\pm4.6)^{\degree} $}& \makecell{$J= s_1s_2s_3c_1^2c_2c_3~\sin\delta$\\$(2.9569\pm0.1330)\times 10^{-5}$} 
\\ \hline

2 & \makecell{ ~~~$\theta_1=0.22623\pm0.00082$ \\~~~$\theta_2=0.00359\pm0.00015$\\~~~$\theta_3=0.04086\pm0.00095$\\ $\delta=(108.4\pm4.5)^{\degree} $}& \makecell{$J= s_1s_2s_3c_1^2c_2c_3~\sin\delta$\\$(2.9569\pm0.1330)\times 10^{-5}$} 
\\ \hline

3 & \makecell{ $\theta_1=0.2261\pm0.0008$ \\~~~$\theta_2=0.00894\pm0.00020$\\$\theta_3=0.0417\pm0.0009$\\ $\delta=(158.1\pm1.0)^{\degree} $}& \makecell{$J= s_1s_2s_3c_1^2c_2c_3~\sin\delta$\\$(2.9569\pm0.1330)\times 10^{-5}$}
\\ \hline

4 & \makecell{ $\theta_1=0.0406\pm0.0009$ \\~~~$\theta_2=0.00350\pm0.00015$\\$\theta_3=0.2262\pm0.0008$\\ $\delta=(107.4\pm4.6)^{\degree} $}& \makecell{$J= s_1s_2s_3c_1^2c_2c_3~\sin\delta$\\$(2.9569\pm0.1330) 10^{-5}$} 
\\ \hline

5 & \makecell{~ $\theta_1=0.0087\pm0.00020$ \\$\theta_2=0.2261\pm0.0008$\\$\theta_3=0.0398\pm0.0009$\\ $\delta=(22.9\pm1.1)^{\degree} $}& \makecell{$J= s_1s_2s_3c_1^2c_2c_3~\sin\delta$\\$(2.9569\pm0.1330)\times 10^{-5}$} 
\\ \hline

6 & \makecell{ $\theta_1=0.0398\pm0.0009$ \\~$\theta_2=0.0087\pm0.00020$\\$\theta_3=0.2264\pm0.0008$\\ $\delta=(157.1\pm1.1)^{\degree} $}& \makecell{$J= s_1s_2s_3c_1^2c_2c_3~\sin\delta$\\$(2.9569\pm0.1330)\times 10^{-5}$} 
\\ \hline

7 & \makecell{ $\theta_1=0.2262\pm0.0008$ \\$\theta_2=0.0156\pm0.0006$\\$\theta_3=0.0389\pm0.0009$\\ $\delta=(94.5\pm4.9)^{\degree} $}& \makecell{$J= s_1^2s_2s_3c_1c_2c_3~\sin\delta$\\$(2.9569\pm0.1330)\times 10^{-5}$} 
\\ \hline

8 & \makecell{ $\theta_1=0.2298\pm0.0008$ \\$\theta_2=0.1756\pm0.0040$\\$\theta_3=0.1792\pm0.0039$\\ $\delta=(178.9\pm0.1)^{\degree} $}& \makecell{$J= s_1^2s_2s_3c_1c_2c_3~\sin\delta$\\$(2.9569\pm0.1330)\times 10^{-5}$} 
\\ \hline

9 & \makecell{ $\theta_1=0.0408\pm0.0009$ \\$\theta_2=0.0858\pm0.0041$\\$\theta_3=0.2156\pm0.0069$\\ $\delta=(93.4\pm4.8)^{\degree} $}& \makecell{$J= s_1^2s_2s_3c_1c_2c_3~\sin\delta$\\$(2.9569\pm0.1330)\times 10^{-5}$} 
\\ \hline
\end{tabular}
\label{analysis}
\end{table} 
 
  Corresponding to the different representations, we have also found the CP violating rephasing invariant  Jarlskog's parameter \textit{J} \cite{jar} defined as 
\be J\sum_{k,\gamma=1}^3 (\epsilon_{ijk}\epsilon_{\alpha\beta\gamma}) = |Im(V_{i\alpha}V_{j\beta}V^*_{i\beta}V^*_{j\alpha})|. \label{j}\ee 
For all the Cartesian representations of the CKM matrix, in column 3 of Table \ref{analysis}, we have presented the corresponding expressions of  
\textit{J}. On numerical evaluation, as expected, its value  comes out to be same for all the representations, also being in agreement with the PDG value \cite{pdg22}, i.e., $(3.08^{+0.15}_{-0.13})\times 10^{-5}$.

Further, for different representations,  we have evaluated  $\epsilon_k$, the CP violation defining parameter in  the $K-\bar{K}$ system. Following Ref.~\cite{review} and using the Cartesian representation 1, this being equivalent to PDG representation,  expressing $V_{cs}$, $V_{cd}$, $V_{ts}$ and $V_{td}$ in terms of the corresponding mixing angles and $\delta$ as well as using the numerical values of these inputs, we get 
 \be \epsilon_k= (2.092 \pm 0.253)\times 10^{-3},\ee this being largely in agreement with the one given by PDG, i.e., $(2.228\pm0.011)\times 10^{-3}$. The same exercise has been carried out for the remaining parametrizations. Intriguingly, out of the 9 representations,  we find that the representations 1, 2, 3, 4, 7 and  8  are able to provide an appropriate fit to the parameter $\epsilon_k$. The other 3 representations, i.e., 5, 6 and 9 are very much off the mark. 
This is not a surprising conclusion keeping in mind the hierarchical nature of the elements of the CKM matrix as well as the accuracy with which these are measured. This has also been discussed in a different context in a recent paper by Xing et. al. \cite{asy3} while evaluating parameter \textit{J} in terms of the magnitudes of the CKM matrix elements.

This also brings to fore whether there is a preferred    representation of the CKM matrix for carrying out phenomenological analyses. To this end, we find PDG representation, perhaps, provides a viable answer to this question. Interestingly, in the PDG representation, within the level of fraction of a percent, the mixing angles capture the hierarchy of the well measured 3 CKM matrix elements. Further, as discussed earlier, the CP violating phase $\delta$ is almost equal to one of the angles of the unitarity triangle. Interestingly, if we compare the PDG representation with the $\gamma$ representation mentioned in equation (\ref{gamma}), one finds that the best measured CKM matrix element $V_{us}$  is expressed in terms of other lesser known elements, unlike the PDG representation. One may also like to compare the PDG representation with the Wolfenstein representation, interestingly, the well known CKM matrix elements as well as the CP violating phase enter indirectly in the   Wolfenstein representation unlike the PDG representation.
  
\section{ Summary and Conclusions}
In the literature, several representations of the CKM matrix have been discussed, however, none of these attempts emphasize clearly the fact that the given representations are the only 9 possible independent ones, nor do these explore the relation between these different representations. Further, keeping in mind the present level of measurement of CKM parameters, these attempts do not explore explicitly the usefulness of a particular representation. In the present work, we have attempted to construct 9 possible independent parametrizations of CKM matrix in rigorous and ab-initio manner, starting with each of the 9 elements of the matrix. The relationship of these independently constructed representations with the already available ones in the literature has been discussed. Further, incorporating unitarity constrains, the implications of these representations have been explored for some of the CKM parameters such as $\delta$, J and $\epsilon_k$. It has been observed that the PDG representation, perhaps, provides the best option for carrying out CKM phenomenology at the present level of measurements. 

\section*{Acknowledgements}
The authors would like to thank the Chairperson, Department of Physics, Panjab University, Chandigarh, for providing the facilities to work.
Gurjit Kaur would also like to acknowledge CSIR, Government of India, Grant No. 09/135/(0851)/2019-EMR-I, for   financial support.


\begin{thebibliography}{99}

\bibitem{ckm1}N. Cabibbo, Phys. Rev. Lett. \textbf{10}, 531 (1963).

\bibitem{ckm2} M. Kobayashi and T. Maskawa, Prog. Theor. Phys. \textbf{49}, 652 (1973).

\bibitem{pdg22} R. L. Workman \textit{et al.} (Particle Data Group), Prog. Theor. Exp. Phys. \textbf{2022}, 083C01 (2022).

\bibitem{ckmfit} Updated results available at http://ckmfitter.in2p3.fr/.

\bibitem{hflav23} Y. Amhis \textit{et al.} (HFLAV Collaboration), Phys. Rev. D \textbf{107}, 052008 (2023).

\bibitem{utfit} A. Bevan \textit{et al.} (UTfit Collaboration), updated results available at http://www.utfit.org/.

\bibitem{jar} C. Jarlskog, Adv. Ser. Direct. High Energy Phys. \textbf{3}, 3 (1989).

\bibitem{fx} H. Fritzsch and Z. Z. Xing, Phys. Rev. D \textbf{57}, 594 (1998).

\bibitem{ar} A. Rasin, arXiv:hep-ph/9708216.

\bibitem{abg} G. N. Li, H.H. Lin, D. Xu and X. G. He, Int. Jour. Mod. Phys. A \textbf{28}, 1350014 (2013).
  
  \bibitem{beta} G. N. Li, H.H. Lin, D. Xu and X. G. He, Phys. Lett. B \textbf{718}, 1454 (2013).

  \bibitem{chau} L. L. Chau and W. Y. Keung, Phys. Lett. \textbf{53}, 1802 (1984).
  
  \bibitem{kmrep} K. Ishikawa, M. Hayashi, T. Morozumi, Soryushiron Kenkyu, \textbf{1}, 2 (2009).

\bibitem{pap} G. Ahuja, M. Gupta, S. Kumar and M. Randhawa, Phys. Lett. B \textbf{647}, 394 (2007).


\bibitem{belle23} M. T. Prim \textit{et al.} (Belle Collaboration), Phys. Rev. D \textbf{108}, 012002  (2023).


\bibitem{nik} N. Awasthi, A. Vashisht, A.  Bagai,  G. Ahuja and M. Gupta, Int. Jour. Mod. Phys. A \textbf{36}, 2150208 (2021).

 \bibitem{asy1} Z.Z. Xing, Phys. Rev. D \textbf{51},3958 (1995).
 
 \bibitem{asy2} Z.Z. Xing, Nuovo Cimento A \textbf{109}, 115 (1996).
 
 \bibitem{asy3} S. Luo and Z. Z. Xing, Nuclear Physics B
\textbf{997}, 116381 (2023).

\bibitem{review} M. Gupta and G. Ahuja, Int. Jour. Mod. Phys. A \textbf{27}, 1230033 (2012).



\end{thebibliography}
\end{document}